\documentstyle[epsfig,aps,preprint]{revtex}
\begin{document}
\draft
\preprint{\vbox{Published in Physical Review {\bf C 51}, 1528 (1995) \hfill}}
\tolerance = 10000
\hfuzz=5pt
\tighten
\title{
Positivity restrictions in polarized coincidence electronuclear scattering}
\author{V. Dmitra\v sinovi\' c}
\address{
Physics Department, University of Colorado, \\
Nuclear Physics Lab, P.O. Box 446, \\
Boulder, CO 80309-0446}
\maketitle
\begin{abstract}
We make a systematic examination of the role played by the restriction that
the cross section in polarized coincidence electronuclear processes must be 
positive. 
The necessary formalism for unpolarized scattering structure functions is 
developed within two frameworks: (i) the response tensor method, and (ii) the 
Jacob-Wick method. Equivalence of the two methods is demonstrated for 
unpolarized scattering, and then the simpler
Jacob-Wick method is applied to the polarized pseudoscalar electroproduction 
off a nucleon. We derive three known and eight new inequalities among
the polarized target structure functions, as well as eleven new polarized 
ejectile structure function inequalities. We also provide rules for this method to be used in the deuteron two-body electrodisintegration in conjunction with 
the results published in Phys. Rev. C {\bf 40} 2479 (1989).
\end{abstract}

\section{Introduction and Summary}

The new and upgraded electron scattering facilities have revived interest 
in coincidence electronuclear scattering \cite{kw83}, and in polarization 
measurements in particular.  
In recent years we have seen a number of theoretical papers 
\cite{dg89,rd89,pvo89} on the general
properties of such cross sections as expressed in three different formalisms. 
A fully relativistic method based on the Jacob-Wick helicity formalism 
\cite{jw59} has been developed for several two-body final state reactions 
\cite{dg89,rtb65,wz68,ddg88,vd90,vd93}. This method allows a comparatively 
straightforward
separation of scattering amplitudes \cite{dg89,ddg88} from the polarization 
observables. Although, at this time complete experimental separation seems 
an excessively ambitious project for all but a few reactions with 
small spins, such theoretical analysis teaches us about the sensitivity of
observables to certain amplitudes, and, as we will show in this paper, it allows
the construction of model-independent inequalities among the observables.
A technically different, but physically related, approach to polarized, 
coincidence, inelastic 
electron scattering was worked out by Donnelly and Raskin \cite{rd89}. They 
use conservation of angular momentum in order to expand the transition 
amplitudes in terms of multipoles, and then write 
the cross section as a function of the latter. While this approach
is closer to the traditional methods of nonrelativistic 
electronuclear physics \cite{dfw66},
it is much more complicated than the helicity amplitude method and it does 
not allow a simple separation analysis. Furthermore, it is not unique insofar
as such multipole analysis can also be completed starting from the Jacob-Wick 
helicity amplitudes \cite{kw83,wz68}.
Finally, a relativistic formalism for the description of target polarization
in coincidence inelastic electron scattering
based on the most general expansion of the response tensor
was developed by De Rujula, Doncel and de Rafael \cite{ddd73}. 
This method, which follows the lines of de Forest's \cite{df67} unpolarized 
coincidence analysis, has recently been extended to describe polarized recoil 
reactions by Picklesimer and van Orden \cite{pvo89}, as well. The final state
channels are not sufficiently well-specified for the separation 
of amplitudes to be feasible within this method.
These three formalisms may look distinct, but they are essentially equivalent:
they all rely on the one-photon-exchange approximation and the use of 
electromagnetic (EM) current conservation.

Despite of all this formal effort, precious few actual calculations 
\cite{pvo89,ar88,mpr93} have been done, so that 
relatively little is known about the actual size, shape, and form of these 
polarized response functions in general. Furthermore, at least one
aspect of the problem remains unexplored: constraints on the polarized 
structure functions due to the condition that the cross section be positive, 
or vanishing (``positive semi-definite").
The {\it inclusive unpolarized} structure functions are well known to be
positive definite \cite{dfw66}. The polarized structure functions, although 
not positive definite, are also bounded by certain functions of the unpolarized ones \cite{don72}, but these bounds are not widely known. As an example of 
potential usefulness, such ``positivity" constraints were used in estimates of 
counting rates for measuring the polarized DIS structure function $g_1$, 
before the experiments.
Some of the above mentioned coincidence scattering methods were developed 
along the lines of inclusive electron scattering, so we expect to find similar 
positivity constraints for coincidence structure functions.

In this paper we present one model-independent set of constraints in the 
form of inequalities among various structure functions based exclusively on 
the positive definiteness of scattering cross sections, {\it i.e.}, 
probabilities. 
Such inequalities have already been derived 
for {\it unpolarized} and several of the {\it polarized target} coincidence 
structure functions \cite{rr73}. One particularly interesting example (see
Sect. II.B below) is the 
upper bound on the absolute value of the so-called ``fifth structure function"
given in terms of the remaining four unpolarized structure functions. 
These results are largely unknown in the nuclear physics 
community, so that they have not seen many applications so far. One of the 
purposes of this paper is to bring some awareness of these results to a wider 
audience. After all, these results are by no means obvious, as are their 
unpolarized inclusive counterparts, and are a natural consequence of the 
probabilistic nature of scattering cross section. 

The method that was originally used \cite{ddd73,rr73} was based on the response 
tensor formalism and turned out to be increasingly complicated 
with the increase of the number of spin degrees of freedom. For that reason,
no double-polarization experiment structure function inequalities are discussed
here.
We develop an alternative scheme based on the Jacob-Wick helicity formalism
that is substantially simpler, especially in view of the fact that all of
the polarization observables that are necessary for this purpose have already 
been worked out for two reactions \cite{dg89,ddg88}. In order to test the
method, we first explicitly show its equivalence with the conventional 
(response tensor) method results for unpolarized coincidence scattering where 
{\it all} inequalities are already known. Then we extend our calculation to the polarized
spin 1/2 ejectile case where we confirm five known inequalities and derive
six new ones. These are by no means all of the inequalities one can
derive in this way. We have only looked at the inequalities that involve up to 
trilinear products of structure functions whereas one can have products of 
up to six structure functions. Rather then spelling out all of them (they 
become increasingly complicated) we indicate how they can be extracted by the 
interested reader from 
the tables provided here, when the need arises. Then we repeat this exercise 
for polarized spin 1/2 target coincidence reactions. Finally, with regard to 
target spin 1 and ejectile spin 1/2 reaction the purpose of this paper is one 
of a do-it-yourself manual. We leave the majority of results to be 
constructed by the interested reader, as the need and occasion arise.
For that purpose, Ref. \cite{vd93} is meant to be used as the source
of tables of polarization observables and their transformation properties. 

This paper is organized as follows. In Section II we review the basics of
the Jacob-Wick and response tensor based formalisms and then present two 
different derivations of the positivity conditions for the {\it unpolarized 
coincidence} structure functions in order to prove their equivalence. In 
Section III we explicitly construct the positivity conditions for spin 1/2 
target and ejectile {\it polarized coincidence} structure functions using the 
second (simpler) of the two methods. We discuss our results and compare them 
with those of Ref. \cite{ddd73}. Then, we define
rules for the derivation of spin 1 polarized target coincidence structure 
functions. 

\section{Review of the Formalism}

In order to set the notational conventions, we review in subsection II.A. 
the basic results of the polarization observables analysis in 
\((e,e^{\prime}N)\) worked out within the Jacob and Wick helicity formalism in 
Refs. \cite{vd90,dg89,vd93}. Then, in subsection II.B. we work out the general 
positivity constraints in unpolarized coincidence electron scattering. 

\subsection{Polarization observables in coincidence electron
scattering}

The general form of the 
inelastic coincidence electron scattering cross section for arbitrary 
polarization of the target and/or ejectile, with two particles in the final 
state, in the ``mixed" frame, {\it i.e.}, with electron variables in the 
laboratory (lab) and the hadronic variables in the centre-of-mass (c.m.) 
frame, is 
\cite{dg89}
\begin{eqnarray} 
{d^{5}\sigma} \over{d\Omega^{\prime} dE^{\prime} 
d\Omega_{1}} & = & 
{\sigma_{M} p_{1} \over 4 \pi M_{T}} 
\bigg\{ \bigg( {W \over M_{T}} \bigg)^{2} v_{L} R_{L} + v_{T} R_{T} 
+ ~ v_{TT} \big[ \cos2\phi~R^{\rm (I)}_{TT} + \sin2\phi~R^{\rm (II)}_{TT}
\big] \nonumber \\
\noalign{\vskip 3pt}
& + &~ \bigg( {W \over M_{T}} \bigg) v_{LT} 
\big[\cos\phi~R^{\rm (I)}_{LT} + \sin\phi~R^{\rm (II)}_{LT}\big] 
\nonumber \\ \noalign{\vskip 3pt}
&+& 2 h ~v_{T}^{\prime} R_{T^{\prime}} 
+ ~ 2 h~\bigg( {W \over M_{T}} \bigg) v^{\prime}_{LT} 
\big[ \cos\phi~R^{\rm (II)}_{LT^{\prime}} + 
\sin\phi~R^{\rm (I)}_{LT^{\prime}}\big] \bigg\} ~, \
\end{eqnarray}
where \(p_1\) is the absolute value of the ejectile three-momentum in the 
c.m. frame and \(E_1\) is the corresponding ejectile energy, 
\(h = \pm 1/2  \) is the helicity of the incoming electron, \(M_T \) is the 
target-nucleus mass, \(\sigma_M \) is the Mott cross-section  
\begin{eqnarray}  
&& 
\sigma_M  = \Big( {\alpha \cos{1 \over 2}\theta \over{2 E \sin^{2}
{1 \over 2}\theta}}\Big)^{2}~, \nonumber
\end{eqnarray}
and
\begin{eqnarray}
v_{L} & = & \bigg( {Q^{2} \over q_{L}^2} \bigg)^2 
~~~~~~~~~~~~~~~~~~~~~~~~~~~~~~~~~~~~ 
v_{TT} = -~{1 \over 2}\bigg( {Q^{2} \over q_{L}^2}\bigg) \nonumber  \\
v_{LT} & = & -~{1 \over \sqrt 2} \bigg({Q^{2} \over q_{L}^{2}} \bigg)
 \sqrt{{Q^{2} \over q_{L}^2} + \tan^{2}{1 \over 2}\theta} 
~~~~~~~~~v_{LT}^{\prime} = -~{1 \over \sqrt 2}\bigg( 
{Q^{2} \over q_{L}^2}\bigg)~ \tan{1 \over 2}\theta \\
v_{T} & = & {1 \over 2} {Q^{2} \over q_{L}^{2}} + \tan^{2}{1 \over 2}\theta
~~~~~~~~~~~~~~~~~~~~~~~~~~~~v_{T}^{\prime} = \tan{1 \over 2}\theta} 
\sqrt{{Q^{2} \over q_{L}^2} + \tan^{2}{1 \over 2}\theta ~. \nonumber \
\end{eqnarray}
The kinematic variables entering 
this cross-section are: the total c.m. energy of the system 
\(W = \sqrt{(P + q)^2} \) (here \(P_{\mu}\) is the target nucleus 
four-momentum), 
the negative four-momentum transfer squared 
\(Q^{2} = -~ q^2 = {\bf q}_{L}^{~2} - \nu^2\), 
the absolute value of the momentum transfer three-vector in the 
lab frame \(q_L = \left|{\bf q}_L \right|\), 
the energy loss \(\nu = E - E^{\prime}\), the initial \(E\) and the final 
state electron energy \(E^{\prime}\) in the lab frame,
the electron scattering angle \(\theta\) in the lab frame,
the ejectile opening angle \(\theta_{1}\) in the c.m. frame and the 
azimuthal angle \(\phi\) (see Fig. 1). We use Bjorken and Drell \cite{bd65} 
metric and \(\alpha \simeq {1 / 137}\) is the fine-structure constant.
The only assumptions entering this result are (i) one-photon 
exchange approximation and (ii) conserved hadron electromagnetic (EM) 
current. Parity conservation has not been assumed as yet. 

The response functions \(R\)'s are functions of \(W, Q^2 \) 
and \(\theta_{1}\), but not of \(\phi\), as long as the target polarization 
is specified with respect to the 
coordinate system \((x^{\prime}, y^{\prime}, z^{\prime})\) (Fig. 1) and  
recoil polarization is measured with respect to the 
\((x^{\prime\prime}, y^{\prime\prime}, z^{\prime\prime})\) 
(Fig. 1) coordinate system. We emphasize this because the specification of the 
recoil polarization measurement coordinate frame has recently been a source 
of some confusion\footnote{It has been erroneously claimed in Ref. \cite{lo91} 
that in the helicity formalism, the recoil polarization is to be measured with 
respect to the 
\((x^{\prime}, y^{\prime}, z^{\prime})\) coordinate system shown in Fig. 1.
An independent calculation of the recoil polarization observables by 
O. Hanstein \cite{han92}, using CGLN amplitudes, 
{\it i.e.} outside of the Jacob-Wick formalism, has shown a discrepancy 
with Lourie's \cite{lo91} results in exactly the form of a rotation through 
angle $\theta_{1}$. This simply and explicitly confirms our claim.}. 
Thus, the ``interference" $R$'s, which do have a $\phi$-dependent factor, can 
be separated by making measurements at different values 
of $\phi$ and otherwise identical kinematics. The remaining ``diagonal" $R$'s
that do not have a $\phi$-dependent factor can be separated by Rosenbluth
separation. The structure 
functions  \(R\)  are linear combinations of functions \(R_{ab} \), which 
in turn, are traces of products of helicity transition matrices 
\(\hat J_{a}, \hat J_{b}^{\dagger}\) (see Eq. (32) of Ref. \cite{dg89}) and the 
density matrices of the initial and the final states: 
\begin{equation}
R_{ab} = (2 s_{1} + 1)(2 s_{2} + 2)~\kappa^{2}~
{\rm tr}\big\{\rho_{f} \hat J_{a} \rho_{i} \hat J_{b}^{\dagger}\big\}~,
\end{equation}
where \(\kappa^2 = {M_{1} M_{2} /{ 4 \pi^{2} W}} \), \(M_{1,2}\), \(s_{1,2}\) 
are the masses and spins of the particles No. 1 (``ejectile") and No. 2 
(``residual nucleus") in the final state, respectively, and all the density 
matrices are normalized to ${\rm tr}(\rho_{f,i}) = 1$. Here the \(a, b\) 
indices stand for the helicities of the virtual photon.
The density matrices can be expanded in terms of irreducible tensor operators 
(ITOs) (see Eqs. (37, 38, 45) of ref. \cite{dg89}) as follows:
\begin{equation}\rho_{i,f}(j) = {1 \over (2 s_{j} + 1)} \sum_{J} \sum_{M} 
T_{JM}^{*}(j) \tau_{JM}(j) \end{equation}
where \(s_{j}\) is the spin value of the \(j^{th}\) particle in the 
initial, or the final 
state, \(\tau_{JM}(j)\) is the \(M^{th}\) component of the spherical ITO 
of rank \(J\), and \( T_{JM}(j)\) are the corresponding components of the 
\(polarization \) spherical ITO. Lorentz boost transformation properties of 
the polarized observables have been worked out in Ref.\cite{vd93}. 
Analogous transformation properties of unpolarized structure functions in 
coincidence electronuclear scattering were worked out by Walecka and Zucker 
\cite{wz68} some time ago. Note that particles No. 2 in their respective
two-body helicity states (the target particle in the initial state, and the 
``residual nucleus" in the final state) carry a tilde above the polarization 
ITO in order to indicate this fact. 

All structure functions can be divided into two
classes: class {\rm I} structure functions ($R_{L},~R_{T}$ besides all other
structure functions explicitly denoted as such by the superscript {\rm I})
are nonzero in $unpolarized,~ parity~ conserving$ reactions, while class
{\rm II} structure functions ($R_{T^{\prime}}$ besides all other structure
functions explicitly marked {\rm II}) vanish identically because of 
constraints imposed by parity. Once the polarization measurement is allowed, 
this rule is modified,
but it remains true that one half of all possible structure functions vanish
because of parity constraints.

\subsection{Unpolarized coincidence positivity inequalities}

The differential cross section $d \sigma$ is a positive or vanishing 
(positive semi-definite), quantity.
Its general form in the one-photon-exchange approximation is proportional to
$l_{\mu}^{*} W^{\mu \nu} l_{\nu}$, where $l_{\mu}$ is the lepton current and
$W^{\mu \nu}$ is the Hermitian positive semi-definite response tensor.
From these two facts alone one can derive nontrivial inequalities among the
structure functions. For inclusive electron scattering the first comprehensive 
study of positivity constraints was published in Ref.\cite{don72}. In the 
following we will
repeat the derivation of the unpolarized structure function inequalities using
two different methods in order to establish their equivalence. The second
method will prove to be much more practical for deriving new, $polarized$
structure function inequalities. Some of the polarized target inequalities have
already been given in Ref.\cite{ddd73}, but all of the polarized ejectile and 
some of the polarized target inequalities presented here are new.

\subsubsection{Positivity inequalities via response tensor method}

Here we extend the original method of Doncel and de Rafael \cite{don72} to 
unpolarized {\it coincidence} scattering structure functions. The derivation 
of the 
positivity inequalities is accomplished by starting from the most {\it general}
unpolarized coincidence cross section for {\it arbitrary} targets and final 
states that is expressed in terms of the so-called response tensor 
$W^{\mu \nu}$: 
\begin{eqnarray} 
W^{\mu\nu} &=& - W_{1} \tilde g^{\mu\nu} +
W_{2}{\tilde p_{T}^{\mu} \tilde p_{T}^{\nu} \over p_{T}^{2}} +
W_{4}{\tilde p_{1}^{\mu} \tilde p_{1}^{\nu} \over p_{1}^{2}} +  \nonumber \\
&+& W_{3}{1 \over {2p_{T} \cdot p_{1}}}
\Big[\tilde p_{T}^{\mu} \tilde p_{1}^{\nu} +
{\tilde p}_{1}^{\mu} {\tilde p}_{T}^{\nu}\Big] + 
W_{5}{i \over {2p_{T} \cdot p_{1}}}
\Big[\tilde p_{T}^{\mu} \tilde p_{1}^{\nu} -
\tilde p_{1}^{\mu} \tilde p_{T}^{\nu}\Big] ~, \
\end{eqnarray} 
where $W_{i}$ ($i= 1-5$) are the five real linearly independent response 
functions and
\begin{eqnarray} 
\tilde g^{\mu\nu} &=& g^{\mu\nu} - {q^{\mu}q^{\nu} \over q^{2}};
~~~\tilde p^{\mu} = p^{\mu} - {p \cdot q \over q^{2}} q^{\mu}~. \nonumber  \
\end{eqnarray} 
The inelastic coincidence electron scattering cross section for 
arbitrary unpolarized target and arbitrary unpolarized ejectile, and any
number of hadrons in the final state, in the lab  frame that follows from 
the above response tensor is \cite{df67}
\begin{eqnarray} 
{d^5 \sigma \over d\Omega^{\prime}dE^{\prime}d\Omega_{L}} &=&
{\sigma_M p_{_L}\over 4\pi M_T} r \biggl\{ v_L R_L + v_T R_T 
+ v_{TT}\left[ \cos 2\phi R^{({\rm I})}_{TT} + \sin 2\phi
R^{({\rm II})}_{TT}\right] \nonumber \\
& +& v_{LT}\left[\cos\phi R^{({\rm I})}_{LT} + \sin\phi R^{({\rm II})}_{LT}
\right] + 2h v_T^{\prime}R_{T^{\prime}}
+ 2h v^{\prime}_{LT}\left[ \cos\phi R^{({\rm II})}_{LT^{\prime}} + \sin\phi
R^{({\rm I})}_{LT^{\prime}}\right]\biggl\}~, \
\end{eqnarray} 
where $d\Omega_{L}, p_{_L}$ are the differential solid angle subtended by, and 
the absolute value of the ejectile lab three-momentum, respectively, and the 
recoil factor $r$ (see Eq. (94) of Ref. \cite{dg89})
$$r = {W \over M_{T}} \left(1 + {{\nu p_{1} - E_{1} q \cos\theta_{1}}\over
{M_{T} p_{1}}}\right)_{L}^{-1} $$
only appears if the final state contains two hadrons. The structure functions 
$R$'s are evaluated in the lab  frame, and are shown in Table II.
This differs from the cross section in the ``mixed" , {\it i.e.}, c.m. and 
lab  frames
through the absence of $W/M_T$ factors in front of the longitudinal structure 
functions (see Eq. (94) of Ref. \cite{dg89}) and a different $\eta$ factor 
in Table I (see below).

We could directly proceed from $l_{\mu}^{*} W^{\mu \nu} l_{\nu} \geq 0$ to find the positivity conditions.
Note, however, that $W_{\mu\nu}$ has only nine independent elements despite 
being a $4 \times 4$ matrix. This, of course, is a consequence
of gauge invariance, which relates the longitudinal and the scalar (zeroth)
components of this tensor. Nevertheless, in general, all 16 matrix elements 
are nonzero. 
The key to simplification, as suggested by Doncel and de Rafael \cite{don72},
is to explictly reduce this four-dimensional matrix to a three-dimensional one 
by a clever choice of reference frame. One solution is to work in the 
Breit frame where the four-vector $q_{\mu}$ loses its temporal component and
becomes
$$q_{B}^{\mu} = (0, {\bf q}_{B}) = (0, 0, 0, Q)~.$$
Then, the current conservation condition
$$q_{\mu} W^{\mu\nu} = q_{\nu} W^{\mu\nu} = 0$$
turns into
$$W^{3,\nu}_{B} = W^{\mu,3}_{B} = 0,$$
{\it i.e.}, only the first three rows and columns survive. It turns out that
this solution can be made frame-independent by introducing the covariant 
photon helicity zero amplitudes. That fact will be proven and used later on,

By finding the relation between the {\it R}'s in the lab frame and the
response tensor $W_{\mu\nu}$ in the Breit frame, we express all of the
elements of $W_{\mu\nu}$  in terms of the observed  lab frame {\it R}'s. 
So, the first step is to work out the boost to the Breit frame. We see that 
the lab$\rightarrow$Breit boost is along 
the $z$ axis so that only the zeroth component is influenced, just like the
lab$\rightarrow$c.m. boost worked out in Ref. \cite{dg89} 
(see Eq.(20) therein). Once again the whole effect of the boost is reduced to
multiplicative factors in front of the L and LT structure functions and a
different phase space factor which is unimportant for this purpose.
To find the multiplicative factor $\eta$ appropriate to the Breit frame we go 
through the same procedure as in Sec. II.B. of Ref. \cite{dg89}.
The Breit frame is defined by
$$q_{B}^{\mu} = \big( 0, 0, 0, q_{B}\big)~, $$
whereas in the lab frame, $p_{1 L}^\mu=(M_{T},{\bf 0})$. The boost
transformation is shown in matrix form as
$$B_{B}=\left(\matrix{{q_{L} \over Q}&0&0&{- \nu \over Q} \cr
0 & 1 & 0 & 0 \cr
0 & 0 & 1 & 0 \cr
{- \nu \over Q} & 0 & 0 & {q_{L} \over Q}} \right)~, $$
so that, if $q_{L}^{\mu}=\big( \nu, 0, 0, q_{L}\big)$, then
$$q_{B} = B_{B} q_{L} = {q_{L} \over Q}q_{L} - {\nu \over Q}\nu = Q~, $$
and we find $\eta = 1$ in the Breit frame, whereas $\eta = {q_{L} \over Q}$ is 
appropriate to the  lab  frame (see Eq.(94) in \cite{dg89}), and  
$\eta = \left({q_{L} \over Q}\right)\left({M_{T} \over W}\right)$ in the c.m.  
frame (see Eq.(29) in \cite{dg89}).
The transverse helicity amplitudes are unchanged by these boosts,
and $\epsilon_{o}^{B}$ in the Breit frame is the time-like unit four-vector
$\epsilon_{o}^{\mu B}=\big(1, {\bf 0}\big) $.
This means that the zeroth component of the response tensor in the Breit frame
is equivalent to the Lorentz-invariant scalar product (contraction) of the 
covariant zero-helicity polarization four-vector and the covariant response 
tensor. This fact will be used later on to define Lorentz-invariant 
inequalities.

The second simplification used is the rotation of the response tensor about
the $z$-axis through $\phi$ which is described in detail in Ref. \cite{dg89}, 
so it will not be dwelt on here. The rotation does not
change the positivity conditions because it is an orthogonal coordinate
transformation. The response tensor defined with respect to the azimuthally 
rotated coordinate system $(x^{'}, y^{'}, z^{'})$ of Fig. 1 will be denoted by
$w_{\mu\nu}$.
The relationship between the observables and the response tensor matrix
elements in the Breit frame is given in Table III. This result allows us to 
invert these relations and express the response tensor matrix elements, 
{\it i.e.}, response functions in terms of observables. 

Now that we have reduced 
the response tensor to a three-dimensional matrix with elements expressed in 
terms of observables, we can apply the mathematical
machinery of positivity constraints on quadratic forms.
The mathematical statement that a quadratic form (matrix) is positive 
semi-definite is equivalent to the statement that  all of its principal 
minors are positive semi-definite, {\it i.e.}, positive or zero. This is easily
proven by diagonalizing the quadratic form and remembering that determinants
(minors) do not change under similarity transformations, {\it i.e.}, under 
rotations of the basis of the linear space.

A direct evaluation of the principal minors of the response tensor yields the
following inequalities (all $w$'s are in the Breit frame and their relation to
the observables is given in Table III):
\begin{mathletters}
\begin{eqnarray} 
w_{00},~w_{xx},~w_{yy} &\geq& 0 \\
\left|\matrix{{w_{00}}&{w_{0x}} \cr
{w_{x0}}&{w_{xx}} \cr}\right| &\geq& 0  \\
\left|\matrix{{w_{00}}&{0} \cr
{0}&{w_{yy}} \cr}\right| &\geq& 0 \\
\left|\matrix{{w_{xx}}&{0} \cr
{0}&{w_{yy}} \cr}\right| &\geq& 0 \\
\left|\matrix{{w_{00}}&{w_{0x}}&{0} \cr
{w_{x0}}&{w_{xx}}&{0} \cr
{0}&{0}&{w_{yy}} \cr}\right| &\geq& 0~, \
\end{eqnarray} 
\end{mathletters}
Note that only inequalities (7a,b) are independent; the remaining two
inequalities (7c,d) being derivable from the first four.
This, together with the hermiticity of the response tensor
$(w_{\mu\nu}=w_{\nu\mu}^{*})$, leads to
inequalities Eq. (8a-c, 10). Specifically, 
from Eq. (7a) and Table III we conclude that
\begin{mathletters}
\begin{eqnarray} 
R_{L} &\geq & 0 \\
R_{T} &\geq & 0 \\
R_{T} &\geq & | R_{TT}^{(\rm I)} |~. \
\end{eqnarray} 
\end{mathletters}
On the other hand, from Eq. (7b) and Table III we obtain:
\begin{eqnarray} 
w_{00} w_{xx} &\geq& | w_{ox} |^{2} \\
4 R_{L} \Big[R_{T} - R_{TT}^{(\rm I)} \Big] &\geq& 
\left( R_{LT}^{(\rm I)} \right)^{2} +
\left(R_{LT^{\prime}}^{(\rm I)}\right)^{2} \
\end{eqnarray} 
which completes our derivation. All of these results hold at each and 
every kinematic point, {\it i.e.},  for every $W^2 , \theta_1 , Q^2 $.

We will re-derive these very same inequalities using a seemingly different 
method and thus establish the equivalence of the methods. This second
method turns out to be much more economical for deriving new, $polarized$
structure function inequalities. Some of the ``polarized target" inequalities
have already been written down in Ref.\cite{rr73}, but all of the polarized 
ejectile and some of the polarized target inequalities presented here are new.

\subsubsection{Positivity inequalities via helicity density matrix method}

In the previous section we have intentionally used 
the helicity states and helicity formalism wherever possible so as to simplify 
the calculation. Specifically,
we have discovered that the linearly independent components of the response
tensor can be rearranged into covariant helicity eigenstates without spoiling 
the 
positivity properties. We build our second method on that simple observation.
It is important to realize that the transition from the Cartesian response 
tensor to the photon helicity states is just an orthogonal coordinate 
transformation that does not change the positivity relations. Furthermore,
by using covariant helicity states of the photon, the whole method becomes 
frame-independent.
Rewriting Eqs.(7a-d) into the helicity (``spherical tensor") basis 
we realize that this is just the positivity constraint for the virtual photon 
density matrix. Density matrices are observables, so they must be Hermitian. 
Moreover, they must be positive semi-definite:
the elements of the diagonalized density matrix are {\it probabilities} of the
system being in the given pure state, and probabilities have to be positive, 
or zero.

Working in the virtual photon helicity basis, we find the 
relationship between the structure functions {\it R}'s in the lab  frame 
and the rotated response tensor matrix elements in the Breit frame that is
shown in Table IV.
In the helicity basis, the response tensor in the ejectile plane is
$$ \rho_{\lambda^{'}\lambda} = 
\varepsilon_{\mu}^{*}(\lambda^{'}) w^{\mu \nu} \varepsilon_{\nu}(\lambda) = 
 \left(\matrix{{w_{++}}&{w_{+0}}&{w_{+-}} \cr
 {w_{0+}}&{w_{00}}&{w_{0-}} \cr
            {w_{-+}}&{w_{-0}}&{w_{--}} \cr}\right)~,  $$
and these inequalities follow from its positive semi-definiteness:
\begin{mathletters}
\begin{eqnarray} 
w_{00},~w_{++},~w_{--} &\geq& 0 \\
\left|\matrix{{w_{++}}&{w_{+0}} \cr
{w_{0+}}&{w_{00}} \cr}\right| &\geq & 0 \\
\left|\matrix{{w_{00}}&{w_{0-}} \cr
{w_{-0}}&{w_{--}} \cr}\right| &\geq & 0 \\
\left|\matrix{{w_{++}}&{w_{+-}} \cr
{w_{-+}}&{w_{--}} \cr}\right| &\geq & 0 \\
\left|\matrix{{w_{++}}&{w_{+0}}&{w_{+-}} \cr
{w_{0+}}&{w_{00}}&{w_{0-}} \cr
{w_{-+}}&{w_{-0}}&{w_{--}} \cr}\right| &\geq & 0 ~,\
\end{eqnarray} 
\end{mathletters}
Using hermiticity and parity conservation (Eqs. (13,15) of Ref. \cite{dg89}), 
we reduce the number of independent real parameters in the virtual photon 
density matrix to five: $w_{00},~w_{++},~w_{+-}$, 
${\rm Re}w_{0+},~{\rm Im} w_{0+}$. Then the inequalities (11b-e) become
\begin{mathletters}
\begin{eqnarray} 
\left|\matrix{{w_{++}}&{w_{0+}^{*}} \cr
{w_{0+}}&{w_{00}} \cr}\right| &\geq & 0 \\
\left|\matrix{{w_{++}}&{w_{+-}} \cr
{w_{+-}^{*}}&{w_{++}} \cr}\right| &\geq & 0 \\
\left|\matrix{{w_{++}}&{w_{0+}^{*}}&{w_{+-}} \cr
{w_{0+}}&{w_{00}}&{-w_{0+}} \cr
{w_{+-}^{*}}&{-w_{0+}^{*}}&{w_{++}} \cr}\right| &\geq & 0 ~,\
\end{eqnarray} 
\end{mathletters}
which leads immediately to the inequalities Eq. (8a,b,c). This proves the
equivalence of the Cartesian general response tensor method, and the
helicity density matrix method for unpolarized structure functions.

This, the second, method is based on the observation that all that is 
necessary is 
an explicit representation of the density matrix in terms of the observables, 
{\it i.e.}, coincidence structure functions. The method by which 
we arrive at this relationship between the density matrix elements and
structure functions is irrelevant. One way is to use a
general response tensor for the reaction
and work out all structure functions and density
matrix elements in terms of the general response functions. This provides
a relation between the density matrix elements and the observables, as desired.
But, that method becomes increasingly complicated as one introduces spin 
\cite{rr73}. Hence, in the future we shall only use the simpler 
helicity amplitude method, but shall compare with the results
of other methods whenever possible.

\subsubsection{Discussion}

The first two inequalities (8a,b) are trivial: they are equivalent to the
statement that the sums of squares of amplitudes have to be positive
or zero. The next
inequality stating that the transverse structure function is larger or equal to
the absolute value of the transverse-transverse interference structure function
is not very surprising, but it was not well-known either. The fourth
result is a non-trivial inequality involving all five structure functions
appearing in the unpolarized cross section, that has not been investigated or
applied to models, so far. It can be used to set bounds on one of the
structure functions if the other four are known. For example, in the so called
impulse approximation the fifth structure function
$R^{({\rm I})}_{LT^{\prime}}$ vanishes identically. We know that this result is unrealistic because all final state interactions are neglected there.
But from the knowledge of $R_{L},~R_{T},~R^{({\rm I})}_{TT},~R^{({\rm I})}_{LT}$ we can obtain an upper bound on the size of $R^{({\rm I})}_{LT^{\prime}}$ in a more realistic approximation:
$$\Big|R^{({\rm I})}_{LT^{\prime}}\Big| \leq
\sqrt{4 R_{L} \Big(R_{T}- R^{({\rm I})}_{TT}\Big)
- \Big(R^{({\rm I})}_{LT}\Big)^{2}}~. $$
These inequalities were first derived in Ref. \cite{rr73},
but they remained largely unknown in the nuclear physics community.
They might prove to be of practical importance as consistency checks of
the experimental extraction procedure for coincidence observables.

The other possibility is to use a
specific reaction whose density matrix can be unambiguously expressed in terms
of its structure functions. This can be done as long
as all possible structure functions are represented, {\it i.e.}, none of them 
vanish ``by accident". The structure
function inequalities obtained in this way are the same as the inequalities
obtained from the general response tensor.

From now on, we will use only
the second, or density matrix, method for the derivation of the polarized
structure function inequalities. Some of the polarized-target inequalities to
be derived have been known for some time (Ref.\cite{rr73}), but many more are 
new.

\section{Polarized Coincidence Structure Function Inequalities}

If another spin degree of freedom, besides that of the photon, is available, 
the density
matrix becomes the direct product of the two density matrices and consequently 
increases in size to $3(2s + 1)\times 3(2s + 1)$, where $s$ is the
(additionally) observed spin. Just as in Sec. II, the elements of
this density matrix are bilinear products of the electromagnetic current induced
transition amplitudes, so that they also can be expressed in terms of the
inelastic electron scattering polarization observables. The
condition for this density matrix to be positive semi-definite leads to
a multitude of inequalities among the polarized structure functions. Once
again, the resulting inequalities are independent of the specific basis vectors
chosen in the Hilbert space. Certain bases, however, are more suitable
for an easy evaluation of the inequalities than others. The substantial 
simplification that comes about due to the use of transversity amplitudes 
\cite{kot66} was observed in Ref. \cite{rr73,ddd73,bds75}. 

The method used to find the relationship between
the density matrix elements and the observables (structure functions) is 
unimportant.
The ``canonical" method \cite{ddd73} is to expand the response tensor in all
possible Lorentz covariants consistent with the general principles, such as the 
current conservation, hermiticity, and parity conservation. Then one expresses 
the structure functions {\it R}'s in terms of the response tensor functions.
After writing the density matrix elements in terms of the response
tensor matrix elements, one can express the density
matrix elements in terms of the observables. Application of the positivity
constraints to the density matrix then automatically leads to inequalities
among the structure functions.

Instead of using this rather cumbersome approach, we choose a specific process
belonging to the given category ({\it i.e.},  having the required spin degrees 
of freedom) and then evaluate its response functions in terms of a finite
number of transition amplitudes.
By expressing the density matrix elements in terms of these very same
transition
amplitudes one gets a one-to-one relationship between the density matrix
elements and the
observables. The only danger in this procedure is in choosing a reaction that
does not allow a unique assignment of observables to the density matrix
elements. An example of such a case would be to take a reaction that has too 
few independent amplitudes, such as the completely spinless reaction, in 
order to determine the unpolarized structure function inequalities. We know 
that, in this case, all amplitudes can be completely separated from 
the five unpolarized structure functions \cite{dg89}.
But, due to this simplicity, one structure function is linearly dependent, 
$R_{T} = \pm R_{TT}$, where the sign depends on the ``parity factor" 
$\eta_g = \pm$ of the reaction, so that one has an ambiguity as to which 
one of the two observables to use in the
density matrix. The solution to this problem is to use a reaction which does
{\it not allow a complete separation of amplitudes from the given set of
observables}. In this way one ensures absence of ambiguities and the results 
one obtains are the same as those obtained by using the general
response tensor method, but with a bigger effort.

In the case of the spin 1/2 target or recoil polarization
measurements, an example of such a ``sample" reaction is scalar or pseudoscalar
electroproduction off a spin 1/2 target. It was proven in Ref. \cite{ddg88} 
that the amplitudes of this reaction cannot be completely separated from only 
one of
these two sets of measurements; thus there is no danger of ambiguity. Secondly,
all of the observables for these two cases have already been worked out in
terms of the transition amplitudes in Ref. \cite{ddg88}. There only remains the task of expressing the density matrices in terms of amplitudes, which is 
accomplished in the next subsection.

\subsection{Polarization observables for spin 1/2 particle target and ejectile 
electroproduction}

We will present the results for $\vec p(e, e^{\prime} \vec p) \pi$ reaction in 
order to illustrate the method. In order to write the density matrix elements 
for $\vec p(e, e^{\prime} \vec p) \pi$ in terms of observables, we 
will use the results of Ref. \cite{ddg88}, as expressed in the notation of
Ref. \cite{dg89}. That allows a simple extension of this method to the
${\vec d}(e, e^{\prime} \vec p) n$ reaction, as well.
Experience has shown that certain linear combinations of helicity amplitudes, 
called hybrid amplitudes, greatly simplify the
resulting tables of observables. These amplitudes are constructed from 
transversity states \cite{kot66}, whose spin quantization axis is 
perpendicular to the reaction plane, for all particles in the reaction except 
the virtual photon. In our 
case this axis is the $y^{\prime} = y^{\prime\prime}$ direction (Fig. 1).
Moreover, the transition from the helicity to the transversity states is
accomplished by a unitary transformation that leaves the positivity properties
of the density matrix intact. The six independent helicity amplitudes are
\begin{eqnarray} 
F_{1}&=<-{1\over 2}\vert J\cdot\epsilon_+\vert -{1\over 2}>
\hskip 30pt
F_{2}=<-{1\over 2}\vert J\cdot\epsilon_+ \vert +{1\over 2}> \nonumber  \\
F_{3}&=<+{1\over 2}\vert J\cdot\epsilon_+\vert -{1\over 2}>
\hskip 30pt
F_{4}=<+{1\over 2}\vert J\cdot\epsilon_+\vert +{1\over 2}> \\
F_{5}&=<+{1\over 2}\vert J\cdot\epsilon_o \vert + {1\over 2}>
\hskip 30pt
F_{6}=<+{1\over 2}\vert J\cdot\epsilon_o \vert - {1\over 2}> ~,\nonumber  \
\end{eqnarray} 
where $J \cdot \epsilon_{\lambda} = J_{\mu} \epsilon^{\mu}_{\lambda}$ and the
initial and final states are helicity states with helicities specified above,
where the nucleon is particle No. 1 in the final state, as defined by Jacob and Wick \cite{jw59}.
$J^{\mu}$ is the hadronic response current defined in the ejectile plane, with
Bjorken and Drell metric \cite{bd65} and
$\varepsilon^{\mu}_{\pm} = (0, {\hat {\bf \varepsilon}}_{\pm}),~
\varepsilon^{\mu}_{0} = (1, {\bf 0}) = (1,0,0,0)$ in the c.m.  frame.

The results are shown in Tables V and VI.
Each structure function appears in the cross section Eq. (1) multiplied by its
spin polarization vector (cf. Eq.(79) of \cite{dg89}) , {\it e.g.}, 
$$R_{LT}^{(\rm I)}= 2 \kappa^{2} \sum_{i,j} p_{i} P_{j}
R_{LT}^{(\rm I)}(p_{i},P_{j})~,$$
where
\begin{mathletters}
\begin{eqnarray} 
R_{LT}(U,U) &=& R_{LT} \\
R_{LT}(U,P_{n}) &=& R_{LT}(n) \\
R_{LT}(p_{x},U) &=& R_{LT}(x)~, \
\end{eqnarray} 
\end{mathletters}
and 
\begin{mathletters}
\begin{eqnarray} 
~a_1~&=&~~{1\over 2} \left(|~g_5~|^{2} + |~g_6~|^{2} \right)~,
~~~a_2~=~{1\over 2}\left(|~g_5~|^{2} - |~g_6~|^{2} \right)\\
~~~b_1~&=&~~~~|~g_1~|^{2}+|~g_2~|^{2}+|~g_3~|^{2}+|~g_4~|^{2}\\
~~~b_2~&=&~~- |~g_1~|^{2}-|~g_2~|^{2}+|~g_3~|^{2}+|~g_4~|^{2}\\
~~~b_3~&=&~~~~|~g_1~|^{2}-|~g_2~|^{2}+|~g_3~|^{2}-|~g_4~|^{2}\\
~~~b_4~&=&~~- |~g_1~|^{2}+|~g_2~|^{2}+|~g_3~|^{2}-|~g_4~|^{2}\\
~~~c_1~&=&~~~~g^*_{4}g_{1}+g^*_3g_{2} ~,
~~~c_2~=~~~~g^*_{4}g_{1}-g^*_3g_{2} \\
~~~d_1~&=&~~~~g^*_{4}g_{5}+g^*_3g_{6}~,
~~~d_2~=~~~~g^*_{4}g_{5}-g^*_3g_{6} \\
~~~e_1~&=&~~~~g^*_{1}g_{5}+g^*_2g_{6} ~,
~~~e_2~=~~~~g^*_{1}g_{5}-g^*_2g_{6} \\
~~~f_1~&=&~~~~g^*_{3}g_{5}+g^*_4g_{6}~,
~~~f_2~=~~~~g^*_{3}g_{5}-g^*_4g_{6} \\ 
~~~k_1~&=&~~~~g^*_{1}g_{3}+g^*_2g_{4}~,
~~~k_2~=~~~~g^*_{1}g_{3}-g^*_2g_{4}~, \
\end{eqnarray} 
\end{mathletters}
We are not interested in complete separation of amplitudes as in Ref. 
\cite{ddg88,dg89}, but rather in separation of certain bilinear products 
that appear in density matrices to be defined below (see Eqs. (17a,b)). 
One can see from Tables V and VI that the moduli squared of the
amplitudes are readily separable from $a_{i},~b_{j}, ~(i=1,2)~(j=1,...,4)$, 
and the same holds for the aforementioned bilinear products. 
An analogous analysis of the $d(e, e^{\prime} N) N^{\prime}$ reaction was
presented in Ref. \cite{dg89}. That provides the necessary information for 
the construction of $s= 1$ polarized target inequalities. 

\subsection{Derivation of polarized positivity inequalities}

In the case of reactions with spin, the 
electromagnetic current matrix elements $J_{a}$ can be assembled into (not 
necessarily square) matrices.
The density matrix with final state spin indices (ejectile polarization)
$$\rho_{\lambda \lambda',tt'}^{f} = 
\left(\matrix{{J_{+} J_{+}^{\dagger}}&{J_{+} J_{0}^{\dagger}}
&{J_{+} J_{-}^{\dagger}} \cr
{J_{0} J_{+}^{\dagger}}&{J_{0} J_{0}^{\dagger}}&{J_{0} J_{-}^{\dagger}} \cr
{J_{-} J_{+}^{\dagger}}&{J_{-} J_{0}^{\dagger}}
&{J_{-} J_{-}^{\dagger}} \cr}\right)  $$
is positive semidefinite \cite{bls80}, just like the initial state density 
matrix (target polarization)
$$\rho_{\lambda' \lambda ,t't}^{i} = 
\left(\matrix{{J_{+}^{\dagger} J_{+}}&{J_{+}^{\dagger} J_{0}}
&{J_{+}^{\dagger} J_{-}} \cr
{J_{0}^{\dagger} J_{+}}&{J_{0}^{\dagger} J_{0}}&{J_{0}^{\dagger} J_{-}} \cr
{J_{-}^{\dagger} J_{+}}&{J_{-}^{\dagger} J_{0}}
&{J_{-}^{\dagger} J_{-}} \cr}\right)  ~.$$
We are free to apply similarity transformations to this density matrix because 
it will not change its positivity
properties. This allows us to change the transverse photon helicity matrix
elements to a new set described below. We are also free to choose any spin 
quantization axis; a particularly convenient one will turn out to be the normal to the ejectile plane. This leads to transversity states \cite{kot66}. We will 
refer to amplitudes $g_{i}$ as the hybrid amplitudes, because the 
photon state is described by its helicity, while all other states are 
transversity states. The longitudinal hybrid amplitudes for pseudoscalar 
electroproduction off an $s = 1/2$ target are defined as follows (for details 
of this construction see Ref. \cite{dg89,ddg88}):
\begin{mathletters}
\begin{eqnarray} 
J_o = \pmatrix{g_{5} & 0 \cr
0 & g_{6}}~. \
\end{eqnarray} 
The transverse hybrid amplitudes can be represented by the following two
matrices:
\begin{eqnarray} 
J_s &=& {1\over 2}(J_{+} - J_{-}) = \pmatrix{0 & g_4 \cr
g_3 & 0} \\
J_a &=& {1\over 2}(J_{+} + J_{-}) = \pmatrix{g_1 & 0 \cr
0 & g_{2}}~.\
\end{eqnarray} 
\end{mathletters}
The specific linear combinations which define the {\it g}'s are
given in Sec. III.A, as are the definitions of helicity amplitudes and the
tables of observables.
It is now straightforward to construct the transformed final state
(recoil polarization) density matrix:
\begin{mathletters}
\begin{eqnarray} 
\rho_{ab,tt'}^f = \pmatrix{
J_a J_{a}^{\dagger} & J_a J_{o}^{\dagger} & J_a J_{s}^{\dagger} \cr
J_o J_{a}^{\dagger} & J_o J_{o}^{\dagger} & J_o J_{s}^{\dagger} \cr
J_a J_{s}^{\dagger} & J_s J_{o}^{\dagger} & J_s J_{s}^{\dagger}}
\end{eqnarray} 
and the initial state (target polarization) density matrix:
\begin{eqnarray} 
\rho_{ba,tt'}^i = \pmatrix{
J_{a}^{\dagger} J_{a} & J_{a}^{\dagger} J_{o}& J_{a}^{\dagger} J_{s} \cr
J_{o}^{\dagger} J_{a} & J_{o}^{\dagger} J_{o} & J_{o}^{\dagger} J_{s} \cr
J_{a}^{\dagger} J_{s} & J_{s}^{\dagger} J_{o} & J_{s}^{\dagger} J_{s}} .
\end{eqnarray} 
\end{mathletters}
The block matrices are
\begin{mathletters}
\begin{eqnarray} 
J_o J_{o}^{\dagger} &=& \pmatrix{|g_{5}|^{2} & 0 \cr
0 & |g_{6}|^{2}} = J_o^{\dagger} J_{o}\\
J_s J_{s}^{\dagger} &=& \pmatrix{|g_{1}|^{2} & 0 \cr
0 & |g_{2}|^{2}} = J_s^{\dagger} J_{s}\\
J_a J_{a}^{\dagger} &=& \pmatrix{|g_{4}|^{2} & 0 \cr
0 & |g_{3}|^{2}} \\
J_a^{\dagger} J_{a} &=& \pmatrix{|g_{3}|^{2} & 0 \cr
0 & |g_{4}|^{2}} \\
J_o J_{s}^{\dagger} &=& \pmatrix{g_{5} g_{1}^{*} & 0 \cr
0 & g_{6} g_{2}^{*}} \\
J_s J_a^{\dagger} &=&  \pmatrix{0 & g_{1}g_{3}^{*} \cr
g_{2}g_{4}^{*} & 0} \\
J_o J_{a}^{\dagger} &=& \pmatrix{0 & g_{5} g_{3}^{*} \cr
g_{6} g_{4}^{*} & 0} \
\end{eqnarray} 
\end{mathletters}
and similarly for other $2\times2$ block matrices that appear in the density 
matrices $\rho_{f,i}$ Eq.~(17a,b). We will express these matrix elements in 
terms of 
observables by using Tables V and VI in Sec. III.A, or  Ref. \cite{ddg88}. 
As an example of derivation, we look at the principal minors of the diagonal 
submatrix $J_o J_{o}^{\dagger}$ Eq.~(17a);
we use Table VI and Eq. (15a,b) to find the two inequalities
\begin{eqnarray} 
R_{L}(U) &\geq& - R_{L}(n) \nonumber  \\
R_{L}(U) &\geq& R_{L}(n) ~,\nonumber  \
\end{eqnarray} 
which immediately lead to 
\begin{eqnarray} 
R_{L}(U) &\geq& |R_{L}(n)| . \nonumber  \
\end{eqnarray} 
Similarly straightforward 
applications of the positivity conditions to other $2\times2$ and $4\times4$ 
diagonal submatrices 
of the initial and final state density matrices lead immediately to 
Eqs. (19,20). Clearly, the larger the submatrix, the more powers of structure 
functions in the inequalities, the highest power in this case being six.

\subsection{Polarized spin 1/2 target and ejectile positivity inequalities}

We start with target polarization inequalities, some of which have been derived
before in Ref.\cite{ddd73}, which results give us another check of the method. 

\subsubsection{Polarized target inequalities}

Proceeding as in Sec. III.B, we find: 
\begin{mathletters}
\begin{eqnarray} 
R_{L}(U) &\geq& |R_{L}(y)| \\
R_{T}(U) &\geq& |R_{T}(y)| \\
R_{T}(U) &\geq& |R^{({\rm I})}_{TT}(y)|  \
\end{eqnarray} 
\begin{eqnarray} 
R_{T}(U) - R^{({\rm I})}_{TT}(U) &\geq&
|R_{T}(y) - R^{({\rm I})}_{TT}(y)| \\
R_{T}(U) + R^{({\rm I})}_{TT}(U) &\geq&
|R_{T}(y) + R^{({\rm I})}_{TT}(y)| ~.\
\end{eqnarray} 
These are the lowest order inequalities that were derived from the ``smallest" 
minors of three diagonal block submatrices. From the next ``larger" minors we 
find the following second order inequalities:
\begin{eqnarray} 
\Big[R_{T}(U) + R_{T}(y)\Big]^{2} &-&
\Big[R^{({\rm I})}_{TT}(U) + R^{({\rm I})}_{TT}(y) \Big]^{2}
 \geq \nonumber  \\
\geq \Big[R_{T^{\prime}}(x) - R^{({\rm II})}_{TT}(z)\Big]^{2} &+&
\Big[R^{({\rm II})}_{TT}(x) + R_{T^{\prime}}(z) \Big]^{2} \\
\Big[R_{T}(U) - R_{T}(y)\Big]^{2} &-&
\Big[R^{({\rm I})}_{TT}(U) - R^{({\rm I})}_{TT}(y) \Big]^{2}
 \geq \nonumber  \\
\geq \Big[R_{T^{\prime}}(x) + R^{({\rm II})}_{TT}(z)\Big]^{2} &+&
\Big[R^{({\rm II})}_{TT}(x) - R_{T^{\prime}}(z) \Big]^{2} \
\end{eqnarray} 
\begin{eqnarray} 
&4& \Big[R_{T}(U) - R^{({\rm I})}_{TT}(U) -
R_{T}(y) + R^{({\rm I})}_{TT}(y)\Big]
\Big(R_{L}(U) - R_{L}(y)\Big) \geq \nonumber  \\
&\geq&\Big[R^{({\rm I})}_{LT}(U) - R^{({\rm I})}_{LT}(y)\Big]^{2} +
\Big[R^{({\rm I})}_{LT^\prime}(U) - R^{({\rm I})}_{LT^\prime}(y)\Big]^{2}\\
&4& \Big[R_{T}(U) - R^{({\rm I})}_{TT}(U) +
R_{T}(y) - R^{({\rm I})}_{TT}(y)\Big]
\Big(R_{L}(U) + R_{L}(y)\Big) \geq \nonumber  \\
&\geq&\Big[R^{({\rm I})}_{LT}(U) + R^{({\rm I})}_{LT}(y)\Big]^{2} +
\Big[R^{({\rm I})}_{LT^\prime}(U) + R^{({\rm I})}_{LT^\prime}(y)\Big]^{2}\\
&4& \Big[R_{T}(U) + R^{({\rm I})}_{TT}(U) +
R_{T}(y) + R^{({\rm I})}_{TT}(y)\Big]
\Big(R_{L}(U) - R_{L}(y)\Big) \geq \nonumber  \\
&\geq&\Big[R^{({\rm II})}_{LT}(x) - R^{({\rm II})}_{LT}(z)\Big]^{2} +
\Big[R^{({\rm II})}_{LT^\prime}(x) + R^{({\rm II})}_{LT^\prime}(z)\Big]^{2}\\
&4& \Big[R_{T^{\prime}}(U) + R^{({\rm I})}_{TT}(U) -
R_{T}(y) - R^{({\rm I})}_{TT}(y)\Big]
\Big(R_{L}(U) + R_{L}(y)\Big) \geq \nonumber  \\
&\geq&\Big[R^{({\rm II})}_{LT}(x) + R^{({\rm II})}_{LT}(z)\Big]^{2} +
\Big[R^{({\rm II})}_{LT^\prime}(x) - R^{({\rm II})}_{LT^\prime}(z)\Big]^{2}\
\end{eqnarray}  
\end{mathletters}
where $x,~y,~z$ stand for target polarization vector components along
the $x^{\prime},~y^{\prime},~z^{\prime}$ directions (Fig. 1), respectively.

Comparison with the results of Ref. \cite{ddd73} is not straightforward for
the following reasons: (i)
some of their inequalities were derived with additional special assumptions
such as that the longitudinal structure function vanish $W_s = \left(1 - 
{\nu^2 \over q^2}\right) W_2 - W_1 = 0$ (Sec. IV of \cite{ddd73}), (ii) the
coordinate frame with respect to which the target polarization is defined is
different ($x \leftrightarrow \pm y$) from ours. We only look at those 
inequalities that can be readily compared. We find that Eqs.~(A 20 a,b,c) of 
\cite{ddd73} correspond exactly to our Eqs.~(19 e,h,i), respectively. 
The two ``trivial Schwarz" inequalities (3.7) of Ref.\cite{ddd73} can be
derived from our Eqs.~(19 f,g) and Eqs.~(19 h-k), but not {\it vice versa}, 
for which reason we do not consider them as independent results. Furthermore,
 De Rujula, Doncel and de Rafael quote three third-order inequalities (their
Eqs.(3.6, A 18, A 19)) which are beyond the scope of this paper.
That leaves us with eight new first- and second-order polarized target 
inequalities. 

\subsubsection{Polarized ejectile inequalities}

Relating the density matrix elements and the
observables and proceeding as in Sec. III.B, we obtain the following
inequalities for recoil polarization structure functions:
\begin{mathletters}
\begin{eqnarray} 
R_{L}(U) &\geq& |R_{L}(n)| \\
R_{T}(U) &\geq& |R_{T}(n)| \\
R_{T}(U) &\geq& |R^{({\rm I})}_{TT}(n)| \
\end{eqnarray} 
\begin{eqnarray} 
R_{T}(U) - R^{({\rm I})}_{TT}(U) &\geq&
|R_{T}(n) - R^{({\rm I})}_{TT}(n)| \\
R_{T}(U) + R^{({\rm I})}_{TT}(U) &\geq&
|R_{T}(n) + R^{({\rm I})}_{TT}(n)| \
\end{eqnarray} 
\begin{eqnarray} 
\Big[R_{T}(U) - R_{T}(n)\Big]^{2} &-&
\Big[R^{({\rm I})}_{TT}(U) - R^{({\rm I})}_{TT}(n) \Big]^{2}
\geq \nonumber  \\
\geq \Big[R_{T^{\prime}}(s) + R^{({\rm II})}_{TT}(l)\Big]^{2} &+&
\Big[R^{({\rm II})}_{TT}(s) - R_{T^{\prime}}(l) \Big]^{2} \\
\Big[R_{T}(U) + R_{T}(n)\Big]^{2} &-&
\Big[R^{({\rm I})}_{TT}(U) + R^{({\rm I})}_{TT}(n) \Big]^{2}
 \geq \nonumber  \\
\geq \Big[R_{T^{\prime}}(s) - R^{({\rm II})}_{TT}(l)\Big]^{2} &+&
\Big[R^{({\rm II})}_{TT}(s) + R_{T^{\prime}}(l) \Big]^{2}
\
\end{eqnarray} 
\begin{eqnarray} 
&4&\Big[R_{T}(U) - R^{({\rm I})}_{TT}(U) -
R_{T}(n) + R^{({\rm I})}_{TT}(n)\Big]
\Big(R_{L}(U) - R_{L}(n)\Big) \geq \nonumber  \\
&\geq&\Big[R^{({\rm I})}_{LT}(U) - R^{({\rm I})}_{LT}(n)\Big]^{2} +
\Big[R^{({\rm I})}_{LT^\prime}(U) - R^{({\rm I})}_{LT^\prime}(n)\Big]^{2}\\
&4& \Big[R_{T}(U) - R^{({\rm I})}_{TT}(U) +
R_{T}(n) - R^{({\rm I})}_{TT}(n)\Big]
\Big(R_{L}(U) + R_{L}(n)\Big) \geq  \nonumber  \\
&\geq&\Big[R^{({\rm I})}_{LT}(U) + R^{({\rm I})}_{LT}(n)\Big]^{2} +
\Big[R^{({\rm I})}_{LT^\prime}(U) + R^{({\rm I})}_{LT^\prime}(n)\Big]^{2}\\
&4& \Big[R_{T}(U) + R^{({\rm I})}_{TT}(U) +
R_{T}(n) + R^{({\rm I})}_{TT}(n)\Big]
\Big(R_{L}(U) + R_{L}(n)\Big) \geq \nonumber  \\
&\geq&\Big[R^{({\rm II})}_{LT^{\prime}}(s) + R^{({\rm II})}_{LT}(l)\Big]^{2} +
\Big[R^{({\rm II})}_{LT}(s) - R^{({\rm II})}_{LT^\prime}(l)\Big]^{2}\\
&4& \Big[R_{T}(U) + R^{({\rm I})}_{TT}(U) -
R_{T}(n) - R^{({\rm I})}_{TT}(n)\Big]
\Big(R_{L}(U) - R_{L}(n)\Big) \geq \nonumber  \\
&\geq&\Big[R^{({\rm II})}_{LT^{\prime}}(s) - R^{({\rm II})}_{LT}(l)\Big]^{2} +
\Big[R^{({\rm II})}_{LT}(s) + R^{({\rm II})}_{LT^\prime}(l)\Big]^{2}~,\
\end{eqnarray} 
\end{mathletters}
where $s,~n,~l$ stand for recoil polarization vector components along the
$x^{\prime\prime},~y^{\prime\prime},~z^{\prime\prime}$ directions (Fig. 1),
respectively.
All of the polarized ejectile results are new, to the best of our knowledge,
and they hold in all $s = 1/2$ ejectile reactions, such as $d(e,e'{\vec p})n$, 
and not just in $p(e,e'{\vec p})\pi$, from which they were derived. 

\subsubsection{Discussion}

Note that these are the inequalities that were obtained from minors
of the two lowest-order diagonal block submatrices. The density matrix is 
$6\times6$ dimensional now, so there are
inequalities with products of up to six response functions ($6^{th}$ order).
Some of these inequalities, however, may turn out to be just products of 
lower order inequalities and hence trivial. 

The higher order inequalities will be useless for
some time to come, because they involve large numbers of spin
observables that are not likely to be measured soon. The high degree of 
nonlinearity makes them quite unlikely to be useful for setting up 
bounds on other observables. For these reasons they are omitted from this work.

Finally, we see that this method can be readily applied to $s = 1$ polarized 
target structure functions as well. The procedure is the same as outlined in
Sec. III.B above: construct the initial state density matrix
$\rho_i$ Eq.~(17b) out of hybrid amplitude matrices $J_{o},J_{a},J_{s}$, shown 
in Eqs( 67,69,70) of Ref. \cite{dg89}, and then set
its principal minors larger or equal to zero. Write the matrix elements of
$\rho_{i}$ in terms of polarized target structure functions given in Tables 
X,~XI~and~XII, to find the desired results. Since polarized deuteron target
measurements are not bound to be made soon, we do not derive such inequalities 
here.

To summarize, in this paper we have
derived three known and 19 new inequalities among the polarized coincidence
inelastic electron scattering structure functions, assuming that the polarized
target, or the polarized ejectile, has spin 1/2. A large number of such 
inequalities have been left for the reader to derive with the help of rules
developed here. This paper completes the development of helicity formalism
as applied to polarized coincidence electron scattering.

\acknowledgments

The author would like to acknowledge useful conversations with O. Hanstein
and would like to thank Prof. R. Brockmann for his hospitality at the 
University of Mainz on two occasions in the summer and fall of 1992, which 
greatly facilitated the inception of this paper. This work was partially 
supported by SURA through a CEBAF fellowship.

\appendix

\section{Relation between helicity and transversity amplitudes }
\label{a:sub}

The connection between the hybrid amplitudes $g_i$
and the helicity amplitudes $F_i$ can be summarized by the matrix relation
$$g_i = \Lambda_{ij}F_j ~,$$
where, for transverse amplitudes ({\it i,j = 1-4}), while for longitudinal 
amplitudes ({\it i,j=5--6}) and
$$\Lambda_{ij}={1\over 2}\left(\matrix{1 & i & -i & 1 & 0 & 0  \cr
1 & -i & i & 1 & 0 & 0 \cr
-i & 1 & 1 & i & 0 & 0 \cr
i & 1 & 1 & -i & 0 & 0 \cr
0 & 0 & 0 & 0 & 2 & -2i \cr
0 & 0 & 0 & 0 & 2 & 2i \cr}\right)~.$$

\widetext
\begin{figure}
\begin{center}
\epsfig{file=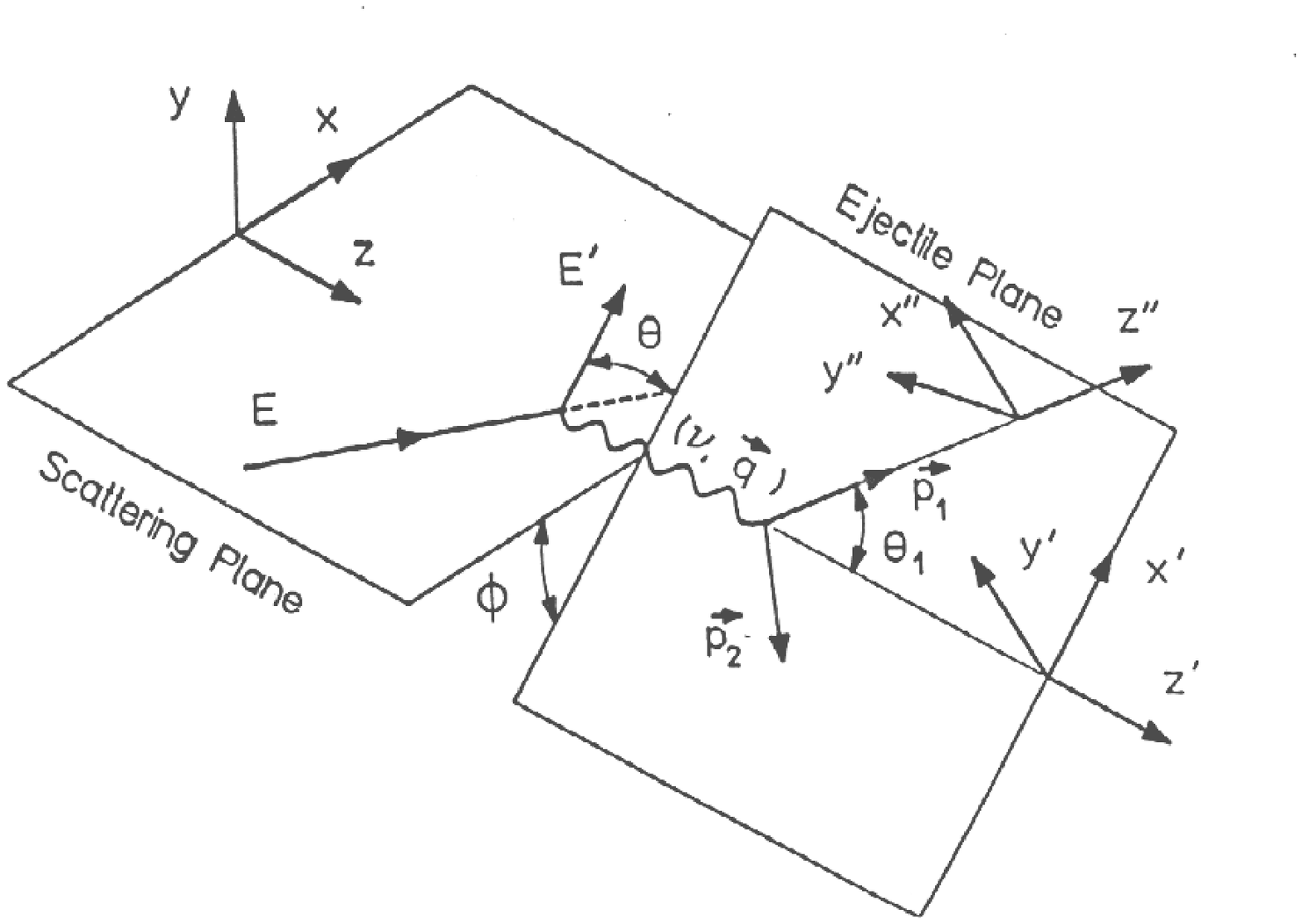,width=16cm} 
\end{center} 
\caption{
Geometry of the coincidence electroproduction or 
electrodisintegration process showing the electron scattering plane, the ejectile plane, and the two coordinate systems in the ejectile plane: \((x^{\prime}, y^{\prime}, z^{\prime})\) defined by the
virtual photon's three momentum ${\bf q}_{L}$ which is parallel to the $z,z'$ 
axes, and \((x^{\prime\prime}, y^{\prime\prime}, z^{\prime\prime})\) 
where $z''$ is along the ejectile three-momentum ${\bf p}_{1}$. The axes 
$y',y''$ are perpendicular to the ejectile plane. The axis $y$ is perpendicular to the electron scattering plane.}
\label{f:geom}
\end{figure}

\begin{table}
\caption{Structure functions $R_{ij}$ expressed in terms of the
response tensor components (second column) or helicity amplitudes (third column), where $\kappa^{2} = {1 \over{4 \pi^{2}}}\left({M_{1} M_{2} \over{2W}}\right)$ 
for both electroproduction and deuteron electrodisintegration 
{\protect\cite{dg89}}.
The helicity amplitudes are defined in the c.m. frame as follows:
$J_{\pm} = J \cdot \epsilon_{\pm} = -
\vec J \cdot \hat \epsilon_{\pm}$ and $J^{o}$ is the zeroth component of the
four-vector $J^{\mu}$ in the c.m.  frame. The subscript 0 in $R_{ij}$'s in the 
second column corresponds to covariant helicity zero states and 
$1/\eta = ({W \over M_T})({Q \over q_L})$ in the c.m.  frame {\protect \cite{dg89}}.}
\begin{tabular}{llll}
\tableline
{$R_L$} & {$ \eta^2 R_{oo}$} & {$  \kappa^2\sum \vert J^o \vert ^2$} \\
{$R_T$} & {$ R_{++}+R_{--}$} & {$ \kappa^2\sum \left(\vert J_+ \vert ^2 +
\vert J_- \vert ^2 \right)$} \\
{$R^{({\rm I})}_{TT}$} & {$ 2 {\rm Re} R_{+-}$} & {$ 2\kappa^2 {\rm Re} \sum (J_+
J_-^{\dag})$} \\
{$R^{({\rm II})}_{TT}$} & {$ - 2{\rm Im} R_{+-}$} & {$ -2\kappa^2 {\rm Im}\sum (J_+
J_-^{\dag})$} \\
{$R^{({\rm I})}_{LT} $} & {$  2\eta {\rm Re} (R_{o+} - R_{o-})$} & {$  2\kappa^2
{\rm Re}\sum J^o \left( J^{\dag}_+ - J^{\dag}_-\right)$} \\
{$R^{({\rm II})}_{LT}$} & {$ 2\eta {\rm Im} (R_{o+} + R_{o-})$} & {$ 2\kappa^2 {\rm Im}
\sum J^o (J^{\dag}_+ + J^{\dag}_- )$} \\
{$R_{T^\prime}$} & {$ R_{++} - R_{--}$} & {$ \kappa^2 \sum (\vert J_+ \vert^2 -
\vert J_- \vert^2)$} \\
{$R^{({\rm II})}_{LT^\prime}$} & {$ 2\eta {\rm Re} (R_{o+} + R_{o-})$} &
{$2\kappa^2 {\rm Re} \sum J^o (J^{\dag}_+ + J^{\dag}_- ) $} \\
{$R^{({\rm I})}_{LT^\prime}$} & {$ 2\eta {\rm Im} (R_{o+} - R_{o-})$} & {$
2\kappa^2 {\rm Im} \sum J^o(J^{\dag}_+ - J^{\dag}_- )$} \\
\end{tabular}
\end{table}
\begin{table}
\caption{ Unpolarized inelastic coincidence structure functions $R$'s in the 
lab frame expressed in terms of the general response tensor functions $W_{i}, 
i = 1 - 5$. Here 
$c = 1 - \left({\nu p_{1} \over{q_{L} E_{1}}}\right)\cos\theta_{1}^{L}$.}
\begin{tabular}{llll}
\tableline
{$R_L$} & {$\Big({q_{L} \over Q}\Big)^{2} \Big \lbrace - W_{1} + \Big({q_{L}
\over Q}\Big)^{2} \Big[W_{2} + c W_{3} + c^{2} W_{4} \Big]\Big \rbrace  $} \\
{$R_T$} & {$2 W_{1} + W_{4} \Big({p_{1} \over M} \sin \theta_{1}\Big)^{2}$}\\
{$R^{({\rm I})}_{TT}$} & {$- W_{4} \Big({p_{1} \over M}
\sin \theta_{1}\Big)^{2} $} \\
{$R^{({\rm I})}_{LT} $} & {$ \sqrt{2} \Big({q_{L} \over Q}\Big)^{2}
\Big({p_{1} \over E_{1}} \sin \theta_{1}\Big)
\Big[W_{3} + 2 c \Big({E_{1} \over M}\Big)^{2} W_{4}\Big] $} \\
{$R^{({\rm I})}_{LT^\prime}$} & {$ \sqrt{2} \Big({q_{L} \over Q}\Big)^{2}
\Big({p_{1} \over E_{1}} \sin \theta_{1}\Big) W_{5} $}
\end{tabular}
\end{table}
\begin{table}
\caption{ Structure functions $R$'s (first column) as functions
of the response tensor components in the helicity basis (second column), or 
as functions of the general response tensor Cartesian components (third column) defined in the Breit frame and in the ejectile plane of Fig. 1; all other 
elements of the general response tensor, e.g. $w_{xy},~w_{0y}$ are zero.
In the Breit frame $\eta = 1$.}
\begin{tabular}{llll}
\tableline
{$R_L$} & {$ \eta^2 R_{oo}$} & {$\eta^{2} w_{00}^{B}$} \\
{$R_T$} & {$ R_{++}+R_{--}$} & {$w_{xx} + w_{yy}$} \\
{$R^{({\rm I})}_{TT}$} & {$ 2 {\rm Re} R_{+-}$} & {$ w_{yy} - w_{xx}$} \\
{$R^{({\rm II})}_{TT}$} & {$ - 2{\rm Im} R_{+-}$} & {$0$} \\
{$R^{({\rm I})}_{LT} $} & {$  2\eta {\rm Re} (R_{o+} - R_{o-})$} &
{$2\sqrt{2}\eta {\rm Re} w_{ox}^{B}$} \\
{$R^{({\rm II})}_{LT}$} & {$ 2\eta {\rm Im} (R_{o+} + R_{o-})$} & {$0$} \\
{$R_{T^\prime}$} & {$ R_{++} - R_{--}$} & {$0$} \\
{$R^{({\rm II})}_{LT^\prime}$} & {$ 2\eta {\rm Re} (R_{o+} + R_{o-})$} &
{$0$} \\
{$R^{({\rm I})}_{LT^\prime}$} & {$ 2\eta {\rm Im} (R_{o+} - R_{o-})$} & {$
2\sqrt{2}\eta {\rm Im} w_{ox}^{B}$} \\
\end{tabular}
\end{table}
\begin{table}
\caption{ Structure functions $R$'s (first column) as functions of the response tensor components in the helicity basis (second column) or as functions of the 
general response tensor components in the helicity basis in the Breit frame and
defined in the ejectile plane (third column). In the Breit frame $\eta = 1$.}
\begin{tabular}{llll}
 \tableline
{$R_L$} & {$ \eta^2 R_{oo}$} & {$\eta^{2} w_{00}^{B}$} \\
{$R_T$} & {$ R_{++}+R_{--}$} & {$2 w_{++}$} \\
{$R^{({\rm I})}_{TT}$} & {$ 2 {\rm Re} R_{+-}$} & {$2 w_{+-}$} \\
{$R^{({\rm II})}_{TT}$} & {$ - 2{\rm Im} R_{+-}$} & {$0$} \\
{$R^{({\rm I})}_{LT} $} & {$  2\eta {\rm Re} (R_{o+} - R_{o-})$} &
{$4\eta {\rm Re} w_{o+}^{B}$} \\
{$R^{({\rm II})}_{LT}$} & {$ 2\eta {\rm Im} (R_{o+} + R_{o-})$} & {$0$} \\
{$R_{T^\prime}$} & {$ R_{++} - R_{--}$} & {$0$} \\
{$R^{({\rm II})}_{LT^\prime}$} & {$ 2\eta {\rm Re} (R_{o+} + R_{o-})$} &
{$0$} \\
{$R^{({\rm I})}_{LT^\prime}$} & {$ 2\eta {\rm Im} (R_{o+} - R_{o-})$} & {$
4\eta {\rm Im} w_{o+}^{B}$} \\
\end{tabular}
\end{table}
\vfil \eject
\begin{table}
\caption{ Recoil polarization observables as functions of
hybrid amplitudes $g_{i}$, where $U$ stands for
unpolarized and $(P_{n}, P_{s}, P_{l})$ are the $(y, x, z)$
components, respectively, of the recoil polarization
vector as measured in c.m.  frame with respect to the
$(x^{\prime\prime}, y^{\prime\prime}, z^{\prime\prime})$ coordinate frame 
(Fig. 1). }
\begin{tabular}{lllll}
 & {~~~U} & {$~~P_n$} & {$~~~P_s$} & {$~~~P_l$}  \\
\tableline
 $R_{L}$ & $~~~a_1$ & $~~a_2$ & {} & {}  \\
 $R_{T}$ & $~~~b_1$ & $~~b_3$ & {} & {}  \\
 $R_{TT}^{(\rm I)}$ & $~~~b_2$ & $~~b_4$ & {} & {}  \\
 $R_{LT}^{(\rm I)}$ & ${\rm Re}(e_1)$ & ${\rm Re}(e_2)$ & & \\
 $R_{LT^{\prime}}^{(\rm I)}$ & ${\rm Im}(e_1)$ & ${\rm Im}(e_2)$ & &\\
 $R_{LT}^{(\rm II)}$ & & & $~~{\rm Im}(d_1)$ & $~~{\rm Re}(d_2)$  \\
 $R_{LT^{\prime}}^{(\rm II)}$ & & & $~~{\rm Re}(d_1)$ & $-{\rm Im}(d_2)$ \\
 $R_{TT}^{(\rm II)}$ & & & $-{\rm Im}(c_1)$ & $~~{\rm Re}(c_2)$ \\
 $R_{T^{\prime}}$ & & & $~~{\rm Re}(c_1)$ & $~~{\rm Im}(c_2)$ \\
\end{tabular}
\end{table}
\begin{table}
\caption{ Same as Table V, but for polarized target observables as functions 
of hybrid amplitudes $g_{i}$, where $U$ stands for
unpolarized and $(p_{x}, p_{y}, p_{z})$ are the $(x, y, z)$
components, respectively, of the target polarization
vector as measured in c.m.  frame with respect to the
$(x^{\prime}, y^{\prime}, z^{\prime})$ coordinate frame (Fig. 1). }
\begin{tabular}{lllll}
 & ~~U & $~~~~p_y$ & $~~~p_x$ & $~~~p_y$  \\
\tableline
  $R_{L}$ & $~~a_1~$ & $~-a_2~$ & &   \\
  $R_{T}$ & $~~b_1~$ & $~~~~b_4~$ & & \\
  $R_{TT}^{(\rm I)}$ & $~~b_2~$ & $~~~~b_3~$ & & \\
  $R_{LT}^{(\rm I)}$ & ${\rm Re}(e_1)$ & $-{\rm Re}(e_2)$ & & \\
  $R_{LT^{\prime}}^{(\rm I)}$ & ${\rm Im}(e_1)~$ & $-{\rm Im}(e_2)$ & & \\
  $R_{LT}^{(\rm II)}$ & & & ${\rm Im}(f_1)$ & $~~{\rm Re}(f_2)$  \\
  $R_{LT^{\prime}}^{(\rm II)}$ & & & ${\rm Re}(f_1)$ & $-{\rm Im}(f_2)$ \\
  $R_{TT}^{(\rm II)}$ & & & ${\rm Im}(k_1)$ & $-{\rm Re}(k_2)$ \\
  $R_{T^{\prime}}$ & & & ${\rm Re}(k_1)$ & $~~{\rm Im}(k_2)~~$ \\
\end{tabular}
\end{table}

\end{document}